\RequirePackage{fix-cm}
\documentclass[smallextended]{svjour3}       % onecolumn (second format)
\smartqed  % flush right qed marks, e.g. at end of proof
\usepackage{graphicx}
\usepackage{amssymb,amsmath}
\usepackage{latexsym}
\usepackage{epstopdf}
\usepackage{color}
\usepackage{epsfig,subfigure}
\begin{document}

\title{Classical dynamics of the Bianchi IX model with timelike singularity\thanks{SP
is grateful to the Ukrainian State Fund For Fundamental Research
for financial support (project F64/45-2016)}
}


\author{Serge L. Parnovsky         \and
        W{\l}odzimierz Piechocki %etc.
}


\institute{S. Parnovsky \at
              Astronomical Observatory, Taras Shevchenko  National University of Kyiv, Observatorna 3a, 04053 Kyiv, Ukraine\\
              Tel.: +380-44-4860021\\
              Fax: +380-44-4862691\\
              \email{par@observ.univ.kiev.ua}           %  \\
           \and
           W. Piechocki \at
              Department
of Fundamental Research, National Centre for Nuclear Research,
Ho{\.z}a 69, 00-681 Warszawa, Poland
}

\date{Received: date / Accepted: date}

\maketitle

\begin{abstract}
We study the dynamics of the vacuum Bianchi IX model with timelike singularity and compare it with the dynamics of the Bianchi IX model
with cosmological singularity. We show that differences in the signs of some terms in the set of equations specifying the dynamics
of both spacetimes lead to significant differences in their properties.
\keywords{Timelike singularities \and Bianchi IX model \and BKL solution}
%\PACS{04.20.Dw \and 04.25.dc \and 02.30.Hq}
\end{abstract}

\section{Introduction}
\label{intro}

Many exact solutions to the Einstein equations describe spacetimes with so-called naked singularities (NS). They have curvature
invariants diverging when approaching the timelike hypersurface of the singularity and have no event horizons, so signals from NS
can reach a distant observer. The existence of at least one NS poses a fundamental problem to modern physics. Energy and information
emanating from NS could affect our Universe and influence its evolution. The so-called Cosmic Censorship Principle was proposed by
Roger Penrose \cite{Pen} to evade this problem\footnote{We recommend the book \cite{PSJ} for comprehensive presentation of this issue.}.
It states that every singularity produced through a collapse must be hidden behind
a horizon, but this is only a hypothesis.

Are NS real objects or just artefact of general relativity? If they were real objects, they could be formed by highly asymmetric collapse
or collapse of rotating, or charged matter. In this case these compact massive objects could be surrounded by accretion disks and
mimic black holes. The recent discovery that the astronomical object named B3 1715+425 could be a nearly naked supermassive
black hole \cite{NNS} increased the interest in searching for NS in the outer space.

A special role in the study of properties of NS is played by the generic timelike singularity proposed in \cite{P1}. It has dynamics of
oscillatory type very similar to the Belinskii, Khalatnikov and Lifshitz (BKL) generic solution  \cite{BKL33} near spacelike
singularity. Both solutions can be matched and form the generic solution near a singularity \cite{P2}. Both conjectures are based on
the generalization of possible solutions to the dynamics of the Bianchi IX model near spacelike and timelike singularities.
These are general analytical considerations based on reasonable assumptions, which are not rigorous mathematically so
BKL solution is not called a theorem, but a conjecture or a scenario. However, due to numerical simulations done in the meantime
(see \cite{Ber,Gar} and references therein), the BKL scenario is commonly believed to underlie the generic general relativity solution.

In this paper we analyse some properties of the solution with NS. We do not emphasize the similarities between both oscillatory
solutions, but differences which completes the paper \cite{P1}. We restrict ourselves, for clarity, to the simplest case of the
vacuum Bianchi IX model.

Recently, two papers appeared with contradictory conclusions verifying the considerations concerning the solution to
the Bianchi IX dynamics near the timelike singularity in terms of the Iwasawa decomposition of the metric \cite{Kli,Sha}.
This was our additional motivation to analyse once again the dynamics of the Bianchi IX spacetime.

Our paper is organized as follows: The next two sections specify the dynamics of the Bianchi IX near spacelike and timelike singularities.
Section IV presents the differences between the dynamics of both cases. We conclude in the last section.

\section{The Bianchi IX model with spacelike singularity} \label{t}

The derivation of all equations has been  presented in \cite{LL,BKL33} so we recall only the main results.
In what follows we use the system of units in which $c = 1$ and the definitions and notations of the book
\cite{LL}. In particular, the signature ($+---$) is used and the Roman indices run over $0,1,2,3$.
In this Section (but not in the Section \ref{x}) the Greek indices $\alpha,\beta,\ldots$ take values
$1,2,3$ and label the spatial coordinates. The coordinate $t=x^0$ is timelike, so we exclude the solutions
describing strong gravitational waves \cite{B}, naked singularities \cite{HP} and other space-times which are
not interesting for cosmology \cite{P79}. The singularity
corresponds to $t=0$.

The metric in a synchronous frame has the form
\begin{equation}
\label{e1}
\rm{d} s^{2} = \rm{d} t^{2} - \gamma_{\alpha \beta}(t)\rm{d} x^{\alpha} \rm{d} x^{\beta} \quad(\alpha,\beta=1,2,3),\\
\gamma_{\alpha \beta}=\eta_{(a)(b)}(t)e_{\alpha}^{(a)}e_{\beta}^{(b)}.
\end{equation}
Here $(a)=1,2,3$ and $\mathbf{e}^{(a)}$ is the set of three frame vectors for the corresponding Bianchi type model.
We use the diagonal tensor $\eta_{(a)(b)}$ because the non-diagonal one is incompatible with the vacuum solution
we consider. Thus, the metric reads
\begin{equation}
\label{e5}
\gamma_{\alpha\beta}=a^2(t) \mathbf{l}_{\alpha}\mathbf{l}_{\beta}+b^2(t) \mathbf{m}_{\alpha}
\mathbf{m}_{\beta}+c^2(t) \mathbf{n}_{\alpha}\mathbf{n}_{\beta},
\end{equation}
where $\mathbf{l},\mathbf{m}$ and $\mathbf{n}$ are the three frame vectors $\mathbf{e}^{(a)}$ of the Bianchi type IX
homogeneous space. They are presented in \cite{LL,Sch} in the form
\begin{equation}
\begin{array}{l}
\label{e5a}
\mathbf{e}^{(1)}=\mathbf{l}=(\sin z,-\cos z\sin x,0),\\
\mathbf{e}^{(2)}=\mathbf{m}=(\cos z,\sin z\sin x,0),\\
\mathbf{e}^{(3)}=\mathbf{n}=(0,\cos x,1).
\end{array}
\end{equation}

All the non-diagonal components of the vacuum Einstein equations  are identically
satisfied, whereas the diagonal components are the solutions to the three equations
\begin{equation}
\label{e6}
2(\ln a)^{\cdot\cdot}=(b^2-c^2)^2-a^4,
\end{equation}
\begin{equation}
\label{e7}
2(\ln b)^{\cdot\cdot}=(a^2-c^2)^2-b^4,
\end{equation}
\begin{equation}
\label{e8}
2(\ln c)^{\cdot\cdot}=(a^2-b^2)^2-c^4,
\end{equation}
and the dynamical constraint
\begin{eqnarray}\label{e9}
\begin{array}{l}
4\big((\ln a)^{\cdot} (\ln b)^{\cdot} +(\ln a)^{\cdot} (\ln c)^{\cdot} +(\ln b)^{\cdot} (\ln c)^{\cdot} \big)\\
= a^4+b^4+c^4 - 2(a^2 b^2 + a^2 c^2 + b^2c^2) \; .
\end{array}
\end{eqnarray}
Here ``dot'' denotes $d/d\tau $, where $\tau$ is the new time variable related with the cosmological time $t$ as follows:
$ d\tau = (abc)^{-1}dt$.

These equations coincides with ones derived for
the Bianchi IX model in Ref. \cite{LL} (see Ch. 14, Sec. 118).
Some analytical studies of the behaviour of the system (\ref{e6}) - (\ref{e9}) have been done in numerous papers (see, e.g.
\cite{Belinski:2014kba} and references therein). Here we make introductory remarks following the paper \cite{BKL}.

The directional scale factors  $a(t),b(t)$ and $c(t)$ make a complex oscillations which could not be described analytically with
all details. An evolution towards the singularity takes a finite interval of the cosmological time $t$, but an infinite interval
of the evolution parameter  $\tau$. An infinite number of oscillations occur during this time interval. They are separated
by the so-called Kasner epochs when the space-time
is similar to the well-known Kasner metric \cite{Kas}
\begin{equation}
\label{eqq14}
\mathrm{d} s^{2} = \mathrm{d} t^{2} - t^{2p_{1}} \mathrm{d} x^{2} - t^{2p_{2}} \mathrm{d} y^{2} - t^{2p_{3}}\mathrm{d} z^{2},
\end{equation}
where the Kasner indices $p_{i}$ satisfy the conditions
\begin{equation}
\label{eqq16}
p_{1} + p_{2} + p_{3} = 1,\quad {p_{1}}^{2}
+{p_{2}}^{2} +{p_{3}}^{2} = 1.
\end{equation}
Thus one of them is negative and other two are positive.  During each Kasner's epoch $\ln (a)$, $\ln (b)$ and $\ln (c)$
are linear functions of $\tau$ and $|\tau|$ increases infinitely as $-\ln (t)$ when approaching singularity.
The system evolves to the singularity if the local volume density $V:= a b c$ decreases, so we get $V\to 0$ as $t\to 0$.
The direction of evolution towards the singularity corresponds to $\tau \to \infty$ or $\tau \to -\infty$ depending on
initial values.
The rule of changing the indices for the adjacent Kasner epochs and
other details is described in \cite{LL,BKL33}.

\section{The Bianchi IX model with timelike singularity} \label{x}

Let us consider the Einstein equations near the
timelike singularity proposed in \cite{P1}. The metric has the form
\begin{equation}
\label{eq1}
ds^2 = -dx^2+\gamma_{\alpha \beta}dx^{\alpha} dx^{\beta},
\end{equation}
where
\begin{equation}
\label{eq5}
\gamma_{\alpha\beta}=a^2(x) \mathbf{l}_{\alpha}\mathbf{l}_{\beta}-b^2(x) \mathbf{m}_{\alpha}
\mathbf{m}_{\beta}-c^2(x) \mathbf{n}_{\alpha}\mathbf{n}_{\beta},
\end{equation}
where $\mathbf{l},\mathbf{m}$ and $\mathbf{n}$ are the three frame vectors of the Bianchi type IX
homogeneous spacetime.

The coordinate $x=x^1$ is spacelike and the singularity corresponds to $x=0$. The other three coordinates
$t=x^0,y=x^2,z=x^3$ use Greek indices, corresponding to 0,2,3. The three frame vectors of the Bianchi type IX
homogeneous space differ from (\ref{e5a}) due to the replacement of one spatial coordinate by $t$.
More specifically, we replace  in (\ref{e5a}) $y$ by $t$, and $x$ by $y$.

All the non-diagonal components of the Einstein equations are identically
satisfied and the diagonal components lead to equations
\begin{equation}
\label{eq6}
2(\ln a)^{\prime\prime}=(b^2-c^2)^2-a^4,
\end{equation}
\begin{equation}
\label{eq7}
2(\ln b)^{\prime\prime}=(a^2+c^2)^2-b^4,
\end{equation}
\begin{equation}
\label{eq8}
2(\ln c)^{\prime\prime}=(a^2+b^2)^2-c^4,
\end{equation}
where ``prime'' denotes $d/d\xi $ and the new coordinate $\xi$ is defined by $d\xi=(abc)^{-1}dx$.
The dynamical constraint has the form
\begin{eqnarray}\label{eq9}
\begin{array}{l}
4\big((\ln a)^{\prime} (\ln b)^{\prime} +(\ln a)^{\prime} (\ln c)^{\prime} +(\ln b)^{\prime} (\ln c)^{\prime} \big)\\
= a^4+b^4+c^4 + 2(a^2 b^2 + a^2 c^2 - b^2c^2) \; .
\end{array}
\end{eqnarray}
For more details concerning the derivation of (\ref{eq6})--(\ref{eq9}) we recommend  \cite{P1}.

\section{Differences in the dynamic of both models} \label{diff}

Two obtained set of equations (\ref{e6}) - (\ref{e9}) and (\ref{eq6}) - (\ref{eq9}) are very similar. Except for differences
in the designation of the variable on which the derivative is taken one must change the sign of the term $a^2$ in the
right-hand side of equations. It looks like Wick rotation. Nevertheless, this small detail provides an essential difference
between solutions. The dynamics of (\ref{e6}) - (\ref{e9}) and (\ref{eq6}) - (\ref{eq9}) are the same if one of the functions
$a,b,c$ prevails over the others and we can neglect the changing of the sign of $a^2$. Thus, the dynamics of the systems are
the same in the main details during most of the evolution. The set (\ref{eq6}) - (\ref{eq9}) also describes an infinite number
of the Kasner epochs. But the set (\ref{eq6}) - (\ref{eq9}) has a singular point at some finite $\xi_0$ and the set
(\ref{e6}) - (\ref{e9}) has not.

This occurs because of the only situation in which the sign of $a^2$ is important, namely the case  $b\approx c\gg a$.
Let us assume that these two functions $b$ and $c$ increase infinitely when approaching the singularity at $\xi=\xi_0$,
while $a$ changes slowly. Thus, we have from (\ref{eq6}) - (\ref{eq8})
\begin{equation}
\label{ez4}
2(\ln (bc))^{\prime\prime}=(a^2+b^2)^2-c^4+(a^2+c^2)^2-b^4 =  2 a^4 + 2a^2 (b^2 + c^2) > 0,
\end{equation}
which means that we are dealing with a convex function.
Let us denote $u=\ln(a^2b^2)$. From (\ref{eq6}) and (\ref{eq7}) we obtain
\begin{equation}
\label{ez6}
u^{\prime\prime}=2c^2(c^2-b^2+a^2).
\end{equation}
Assume that
\begin{equation}
\label{ez6a}
|b^2-c^2|\ll a^2\ll b^2.
\end{equation}
This equation simplifies to
\begin{equation}
\label{ez7}
u^{\prime\prime}=2e^u.
\end{equation}
There are three types of solutions of (\ref{ez7}), namely
\begin{equation}
\label{ez8}
e^u=a^2b^2=(\Delta\xi)^{-2},
\end{equation}
\begin{equation}
\label{ez9}
e^u=a^2b^2=p^2\sin^{-2}(p\Delta\xi),
\end{equation}
\begin{equation}
\label{ez10}
e^u=a^2b^2=p^2\sinh^{-2}(p\Delta\xi).
\end{equation}
Here $\Delta\xi=\xi-\xi_0$ and $\xi_0$, $p$ are constants. All these solutions have singularities at some finite value $\xi=\xi_0$
with $a^2b^2\xrightarrow [\xi \to \xi_0]{}(\Delta\xi)^{-2}$.

The equation (\ref{eq6}), in the case  (\ref{ez6a}), leads to
\begin{equation}
\label{ez11}
2(\ln a)^{\prime\prime}=-a^4 ,
\end{equation}
which has the solution
\begin{equation}
\label{ez11}
a=S^2 \cosh^{-2}(S(\xi-\xi_1))~~~ S,\xi_1=\mathrm{const}.
\end{equation}
So the function $a$ is regular and $b^2\approx c^2\to G\;(\Delta\xi)^{-2}$
at $\xi \to \xi_0$. Here $G=\mathrm{const}$.

From (\ref{eq7}) and (\ref{eq9}) we get
\begin{equation}
\label{ez5}
2(\ln (b/c))^{\prime\prime}=2a^2(c^2-b^2)+2c^4-2b^4=2(c^2-b^2)(a^2+b^2+c^2).
\end{equation}
Let us denote $w=\ln(b^2/c^2)$ and rewrite (\ref{ez5}) for the case (\ref{ez6a})
and $|w|\ll 1$ in the form
\begin{equation}
\label{ez12}
w^{\prime\prime}=-4wc^4=-4G^2w(\xi-\xi_0)^{-4}.
\end{equation}
Its solution is
\begin{equation}
\label{ez13}
w=\Delta\xi\left(C_1\sin\frac{2G}{\Delta\xi}+C_2\cos\frac{2G}{\Delta\xi}\right),\quad C_1,C_2=\mathrm{const}.
\end{equation}
So the condition (\ref{ez6a}) can be fulfilled near the singularity.
At small $\Delta\xi$ both sides of the constraint (\ref{eq9}) lead to $4(\Delta \xi)^{-2}$.
We see that the constraint is satisfied in the leading terms.

We will analyse it later. Now we will prove that the similar solution is impossible in the case of the set
(\ref{e6}) - (\ref{e9}).
If two of the functions, for example $a$
and $b$, increase when approaching the singularity, while the other one changes slowly, i.e. we have the case $a\approx b \gg c$.
Thus, we have
\begin{equation}\label{ez}
2(\ln (ab))^{\cdot\cdot}= (b^2 - c^2)^2 - a^4 + (a^2 - c^2)^2 - b^4 = 2 c^4 - 2 c^2 (a^2 + b^2) < 0.
\end{equation}
Thus, this is a concave function. This fact contradicts our assumption. So, the is no solution of the set (\ref{e6}) - (\ref{e9})
similar to considered above one with singular point at $\tau=\tau_0$. There is no other solution of sets (\ref{e6}) - (\ref{e9})
and (\ref{eq6}) - (\ref{eq9}) with singularity at finite $\xi$ or $\tau$ with other $a,b,c$ asymptotic relation (see the
analysis in \cite{LL}).

 %%%%%%%%%%%%%%%%%%%%%%%%%%%%%%%%%%%%%%%%%%%%%%%%%%%%%%%%%%%
%    Fig.1
%%%%%%%%%%%%%%%%%%%%%%%%%%%%%%%%%%%%%%%%%%%%%%%%%%%%%%%%%%%
\begin{figure}[tb]
\includegraphics[width=\columnwidth]{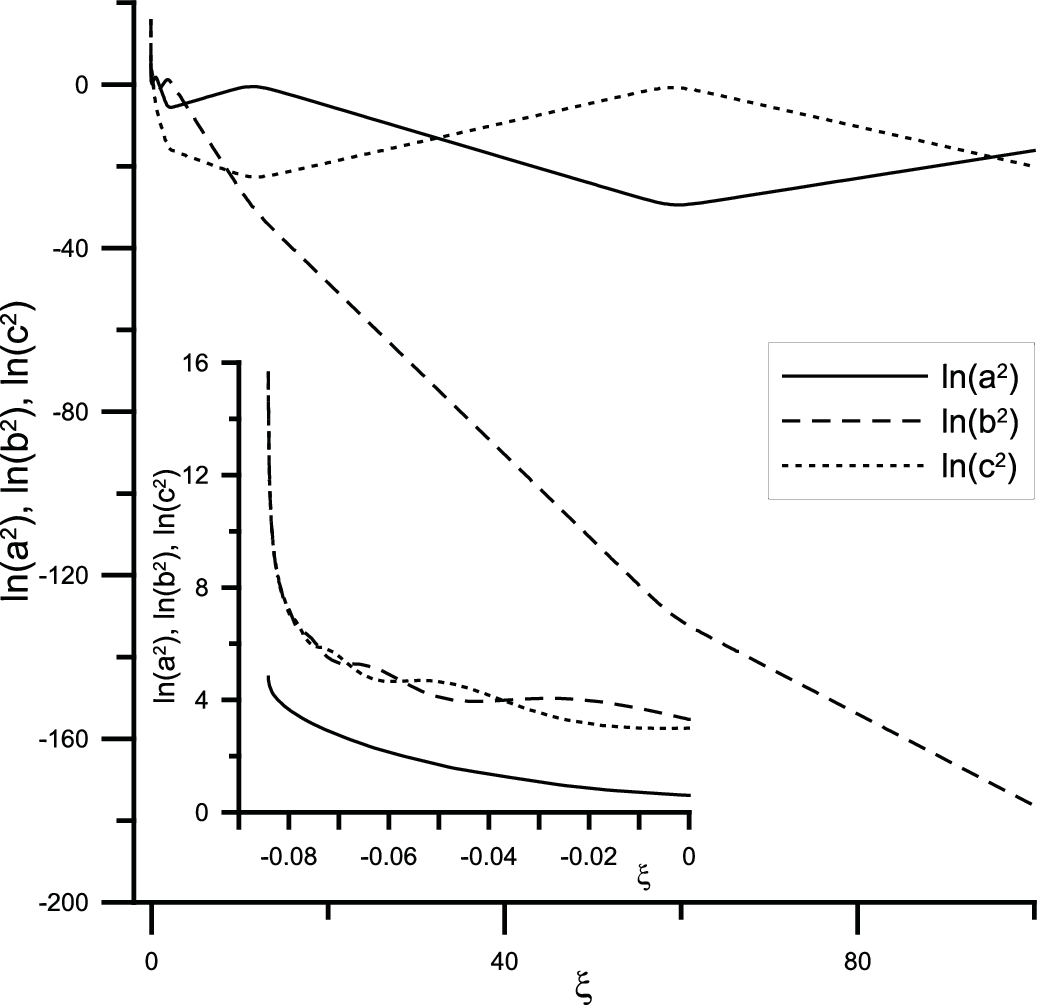}
\caption{ An example of solution to the system of equations (\ref{eq6}) - (\ref{eq8})
with initial values and derivatives satisfying the condition (\ref{eq9}). There is a
singularity with $b,c \to \infty$ at small negative $\xi_0$}
\label{f2}
\end{figure}
%%%%%%%%%%%%%%%%%%%%%%%%%%%%%%%%%%%%%%%%%%%%%%%%%%%%%%%%%%%
%%%%%%%%%%%%%%%%%%%%%%%%%%%%%%%%%%%%%%%%%%%%%%%%%%%%%%%%%%%
%    Fig.2
%%%%%%%%%%%%%%%%%%%%%%%%%%%%%%%%%%%%%%%%%%%%%%%%%%%%%%%%%%%
\begin{figure}[tb]
\includegraphics[width=\columnwidth]{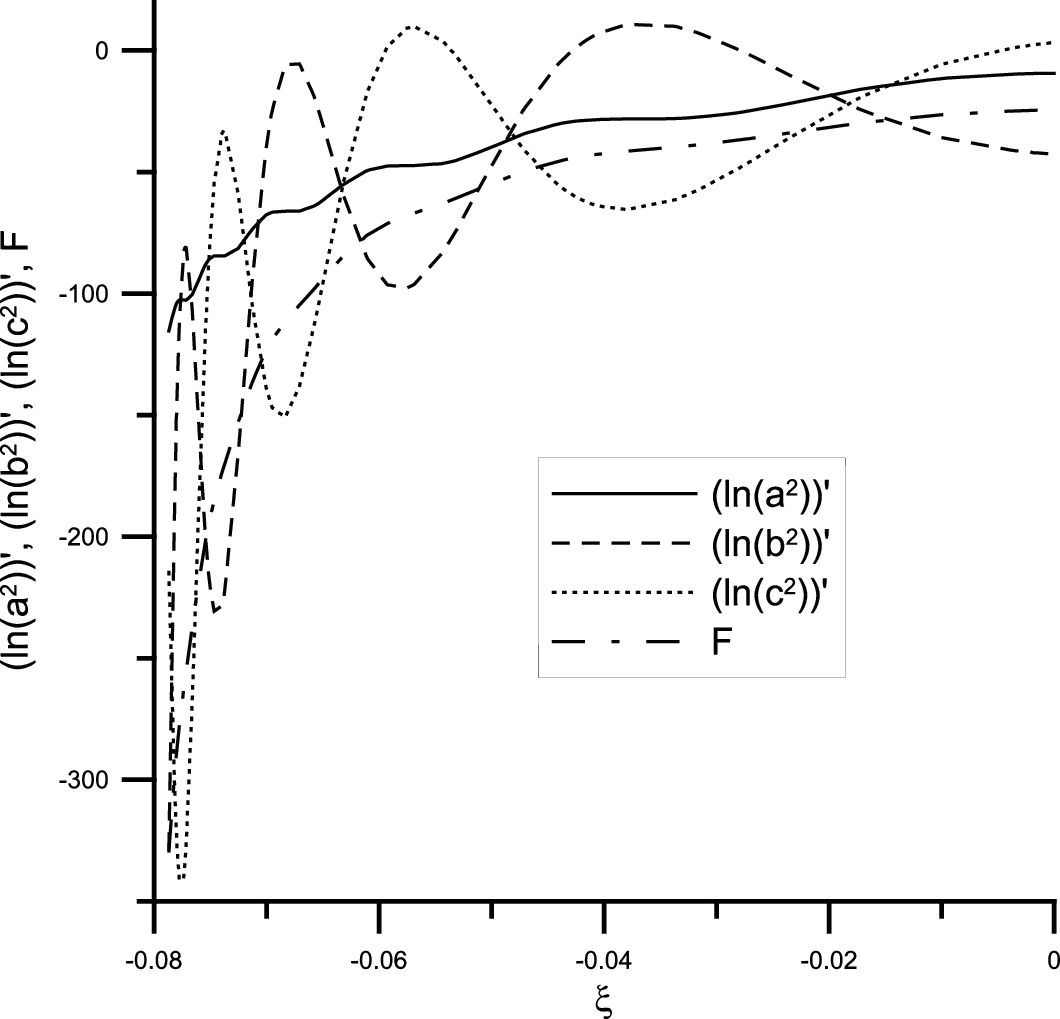}
\caption{ Derivatives $(\ln a)^{\prime},
(\ln b)^{\prime},(\ln c)^{\prime}$ and $F$ vs $\xi$ at $\xi<0$}
\label{f4}
\end{figure}
%%%%%%%%%%%%%%%%%%%%%%%%%%%%%%%%%%%%%%%%%%%%%%%%%%%%%%%%%%%

One can see an example of such singularity in Fig. \ref{f2}. It presents the numerical solution
of equations (\ref{eq6}) - (\ref{eq8}) with initial values and derivatives with respect to $\xi$ satisfying
(\ref{eq9}). It illustrates the above-mentioned solution together with Fig. \ref{f4} which shows the dynamic of derivatives
$(\ln a)^{\prime}$, $(\ln b)^{\prime}$ and $(\ln c)^{\prime}$ near the singularity, as well as their sum
\begin{equation}\label{f}
F=\mathrm{d}\ln(abc)/\mathrm{d}\xi=\mathrm{d}\ln(V)/\mathrm{d}\xi,
\end{equation}
where $V$ is the local volume. The sign of $F$ tell us what value of $\xi$ corresponds to singularity $x=0$.
In Fig. \ref{f2} we have $F<0$ at $\xi=0$. Increasing $\xi$ we move towards an oscillating timelike
singularity, whereas decreasing $\xi$ we move towards a singularity with $b,c \to \infty$. In the case
$F>0$, at $\xi=0$, we have an oscillating solution as $\xi \to -\infty$ and the discontinuity
at some positive $\xi_0$. In both cases $|F|$ decreases if we move towards the oscillating timelike singularity
corresponding to $F=0$.

The function $F$ is useful when we consider the possibility that the required solution is defined
on the interval ($\xi_1,\xi_2$) with two singularities at $\xi=\xi_1$ and $\xi=\xi_2$. It requires a change of the sign
of $F$ and is ruled out by the fact
that this sign is conserved. Indeed, summing up (\ref{eq6}) - (\ref{eq9}) we get
\begin{equation}
\label{ez14}
2F^{\prime}=(b^2-c^2)^2+a^4+2a^2(b^2+c^2)>0.
\end{equation}
Assume that $F$ becomes zero at some $\xi$. In this case we have
$(\ln(a))^{\prime}=-(\ln(b))^{\prime}-(\ln(c))^{\prime}$ and the left-hand side of (\ref{eq9}) is
$$4\left(-\left((\ln b)^{\prime}\right)^2 -\left((\ln c)^{\prime}\right)^2+(\ln b)^{\prime} (\ln c)^{\prime}\right)<0,$$
while its right-hand side is $a^4+(b^2-c^2)^2 + 2a^2( b^2 + c^2)>0$. This contradiction proves that
the function $F$ keeps its sign and eliminates the possibility of $F(\xi)$ having two  discontinuities at
different values of $\xi$. Note that the analogous function for cosmological model also keeps its sign, although due to another reason.
So, the local volume $V$ increases monotonically with increasing distance from the singularity at $x=0$.

Is it naked? Let us move towards it along the spacelike $x$ coordinate. Nevertheless, if there was a horizon around the singularity
it could looks like coordinate singularity with finite curvature invariants. For example, the Schwarzshild horizon in the
synchronous frame looks like generalized Kasner metric with indices (1,0,0) i.e. the false singularity.
Let us study the singularity at $\xi=\xi_0$. It corresponds not to horizon, but to infinitely distant hypersurface.
Indeed, the distance along the line
with constant $y,z,t$ between two points in spacetime (\ref{eq1}) is equal to the difference of their $x$ coordinates
or $\Delta x=\int_{x_1}^{x_2}abc\, d\xi$. Such a distance between the point $\xi=0$ and the oscillating
singularity at $\xi=\pm \infty$ in Fig. \ref{f2} is finite and a distance between the $\xi=0$ point and the singularity at $\xi=\xi_0$
is infinite.

So the singularity at finite $\xi$ is located infinitely far from the oscillating timelike singularity and corresponds to $x=\infty$.
This is not a real singularity, but just a coordinate one. The asymptotic form of the metric (\ref{e5}) near this
false singularity at $x\to \infty$ is
\begin{eqnarray}\label{ez15}
\begin{array}{l}
\displaystyle \rm{d} s^{2} \xrightarrow [x \to \infty]{} -\rm{d} x^{2} +C_1(\cos y
\, \rm{d}t+\rm{d}z)^2
-C_2 x^2(\cos z\, \rm{d}y +\sin y \sin z\, \rm{d}t)^2\\
\displaystyle  -C_2\left(x^2+C_3\frac{\sin(C_4x)}{x}\right)(\sin z \,\rm{d}y -\sin y \cos z\, \rm{d}t)^2
\end{array}
\end{eqnarray}
with $C_i=$ const.

The calculations of both the Ricci and the Kretschmann scalars with the metric  (\ref{ez15}) give the result that they both
vanish at  $x \to \infty$. Thus, the approximate metric (\ref{ez15}) is asymptotically flat and it may describe the metric  near
the remote edge of considered spacetime infinitely far away from the real timelike oscillatory-type singularity that occurs as
$x \to 0$. We see that the singularity at $\xi=\xi_0$ is not a horizon, so the singularity at $x=0$ of space-time described by the metric
(\ref{eq1}) is naked one.

Could there be a solution of the equations (\ref{eq6}) - (\ref{eq9}) without coordinate singularity (\ref{ez15}) at finite value of $\xi$?
Such solution was determined for all $\xi$ values from $-\infty$ to $\infty$ like the BKL one is determined for all $\tau$ values.
It seems to us that this is not possible, although we can not prove this assumption. Let us express some arguments in
favour of this conclusion. Initial parameter space for (\ref{eq6}) - (\ref{eq8}) is six-dimensional because it describes the initial
values of three functions $a,b,c$ and three derivatives $a^{\prime},b^{\prime},c^{\prime}$. The condition (\ref{eq9}) fixes the
five-dimensional region of possible initial parameters. The additional symmetry with respect to the transformation
\begin{equation}\label{ez1}
\tilde{a}=Ka,~~~\tilde{b}=Kb,~~~\tilde{c}=Kc,~~~\tilde{\tau}=K^{-2}\tau,~~~K=const.
\end{equation}
effectively makes it a four-dimensional one. We can add the $\xi$ coordinate and consider the dynamics of the set of
equations (\ref{eq6}) - (\ref{eq8}) in the seven-dimensional space, more precisely in the five-dimensional subspace. Trajectories, which
start from some initial points or sets of initial parameters end, as a rule, in singularities of type (\ref{ez15}).

The problem is whether there are trajectories which avoid singularities at finite $\xi$. If such trajectories exist, a second
question arises. What should be the dimension of a set of starting points with trajectories avoiding such singularities, i.e. trajectories reaching
$\xi=\infty$ or $\xi=-\infty$ far from singularity at $x=0$. These problems are complex from a mathematical point of view. We
present some speculations instead of a comprehensive solution.

Remember that the dynamical evolution consists of a finite or infinite set of oscillations. After any oscillation till the next one
we have a ``world line'' in the seven-dimensional space with coordinates $\xi,a,b,c,a^{\prime},b^{\prime},c^{\prime}$.
If the coordinates $a,b,c,a^{\prime},b^{\prime},c^{\prime}$ in the six-dimensional parameter space reach a region where
condition (\ref{ez6a}) and some restrictions on $a^{\prime},b^{\prime},c^{\prime}$ are fulfilled, then the singularity of
type ({\ref{ez15}) is inevitable. If BKL oscillation had some hidden invariants, the requirement for their conservation could
prevent the system from getting into these areas. However, oscillations in the Bianchi IX space are considered to be chaotic
\cite{BBE,HBC,R,DDR}.
If there is an arbitrarily small but nonzero probability that the system after some oscillation will be in the region of
the parameters space where the singularity is inevitable, then sooner or later it will be reached.
Even a vanishingly small probability has to be realized in an infinite number of attempts.

\section{Conclusions}
\label{conc}

We have examined the properties of the spacetime defined by the metric (\ref{eq1}) which is homogeneous
on the hypersurface $x=$ const.
This homogeneous cross-section corresponds to the Bianchi type IX case. There is a naked timelike singularity at
$x=0$. This model is the basis for constructing the general solution near the timelike singularity \cite{P1}
of oscillating type, which is similar to the well-known spacelike singularity \cite{BKL}.

The dynamics of the model is similar to the dynamics of the vacuum homogeneous cosmological model of Bianchi type IX with
the change of coordinates $x\leftrightarrow t$.
In both cases we have oscillations of directional scale factors in the form of a sequence of eras
with each era consisting of a number of epochs.
Nevertheless, there are some differences between the model with the metric defined by (\ref{eq1})
and the cosmological model with the metric defined by (\ref{e1}). Let us mention the main differences: The coordinate
$\tau=\int \gamma^{-1/2} dt$ with spacelike singularity at $t=0$ varies from $-\infty$ to $\infty$.
The coordinate $\xi=\int \gamma^{-1/2} dx$ with timelike singularity at $x=0$ varies from $\pm \infty$
to some finite value $\xi_0$. The case $\xi \to \pm \infty$ corresponds to the naked timelike oscillating singularity.
The singularity at $\xi=\xi_0$ corresponds to a infinitely remote point; it is a false one as the curvature invariants vanish there.
An asymptotic form of metric is described by (\ref{ez15}).

If one started from any point in space and moved along the $x$ coordinate to the point
infinitely far from the naked singularity, he or she would observe that spacetime makes a finite number of BKL-like oscillations and
an infinite number of oscillations of (\ref{ez15}) type. In the case of travel along the time coordinate $t$ away from
the cosmological spacelike singularity the number of BKL-like oscillations is infinite and there are no oscillations like (\ref{ez15}).
In both cases the evolution towards the singularity includes an infinite number of BKL-like oscillations.

Our results are consistent with the recently obtained results for the diagonal Bianchi IX model with timelike
singularity, which are based on the Iwasawa decomposition of the metric \cite{Sha}. The difference between the results of
Refs. \cite{Sha} and \cite{Kli},  which uses Iwasawa's decomposition as well, might be due to the quite  different choices of gauges.
However, deeper analysis explaining this inconsistency is beyond the scope of the present paper.

We prove that the sign of $F=\mathrm{d}\ln(V)/\mathrm{d}\xi$
as well as $\mathrm{d}V/\mathrm{d}x$ is constant. Therefore, the local volume $V$ increases monotonically with increasing
distance from the naked singularity at $x=0$. This fact eliminates the possibility of the existence of several singularities
with finite values of $\xi$. This is especially important when analyzing the dynamics of spacetime without analytical
expressions for $a(x),b(x),c(x)$.

\begin{acknowledgements}
We are grateful to Orest Hrycyna for his numerical calculations for the purpose of Figs. (\ref{f2}) -
(\ref{f4}), and Grzegorz Plewa for the calculations concerning the curvature invariants. We also thank John Barrow, Vladimir Belinski,
Piotr Chru\'{s}ciel, and Edgar Shaghoulian for helpful discussions.
\end{acknowledgements}

\end{document}